\documentclass[a4paper,aps,superscriptaddress,floatfix,nofootinbib,twocolumn,longbibliography]{revtex4-1}

\usepackage{amsmath, amsthm, amssymb} 
\usepackage{algpseudocode} 
\usepackage{comment} 
\usepackage{graphicx}
\usepackage{dcolumn}
\usepackage{bm}
\usepackage{hyperref}
\usepackage[mathlines]{lineno}
\usepackage{float}
\usepackage{booktabs}
\usepackage{multirow}
\usepackage{xcolor} 

\theoremstyle{definition}

\AtBeginDocument{
	\heavyrulewidth=.08em
	\lightrulewidth=.05em
	\cmidrulewidth=.03em
	\belowrulesep=.65ex
	\belowbottomsep=0pt
	\aboverulesep=.4ex
	\abovetopsep=0pt
	\cmidrulesep=\doublerulesep
	\cmidrulekern=.5em
	\defaultaddspace=.5em
}

\begin{document}

\title{A typology of activities over a century of urban growth}


\author{Julie Gravier}
\affiliation{Centre de Recherches Historiques, EHESS (CNRS/EHESS) 54 Avenue de Raspail, 75006 Paris, France}
\affiliation{Centre d'Analyse et de Mathématique Sociale (CNRS/EHESS) 54 Avenue de Raspail, 75006 Paris, France}
\author{Marc Barthelemy$^{* ,2,}$}
\affiliation{Universit\'e Paris-Saclay, CNRS, CEA, Institut de Physique Th\'eorique, 91191, Gif-sur-Yvette, France
 $^*$marc.barthelemy@ipht.fr}

\begin{abstract}

Contemporary literature on the dynamics of economic activities in growing cities mainly focused on a few years or decades time frames. Using a new geo-historical database constructed from historical directories with about 1 million entries, we present a comprehensive analysis of the dynamics of activities in a major city, Paris, over almost a century (1829-1907). Our analysis suggests that activities that accompany city growth can be classified in different categories according to their dynamics and their scaling with population: (i) linear for everyday needs of residents (food stores, clothing retailers, health care practitioners), (ii) sublinear for public services (legal, administrative, educational), (iii) superlinear for the city's specific features (passing fads, specialization, timely needs). The dynamics of these activities is in addition very sensitive to historical perturbations such as large scale public works or political conflicts. These results shed light on the evolution of activities, a crucial component of growing cities.

\end{abstract}

\maketitle


\section*{Introduction}

The science of cities benefited these last years from the recent availability of massive amount of data about various aspects of these systems~\cite{batty2013new,barthelemy2016structure,bettencourt2021introduction} such as mobility, segregation, $\mathrm{CO}_2$ emissions, etc. The growth dynamics is one of the most important aspect as it results from the evolution of infrastructures, the population distribution and economic activity. A large amount of studies discussed the growth of cities and their main drivers~\cite{duranton2014growth}, highlighting the importance of innovation \cite{pumain_evolutionary_2006}, amenities, agglomeration economies and human capital. Political institutions, the degree of democratization and technological advances also naturally play a key role~\cite{Henderson}. Shocks are also very important and determine the fate of cities. These shocks can be structural changes or related to new industries and affect the urban landscape and eventually have a large impact on interurban migration~\cite{Verbavatz}.

Most of the available datasets for this type of studies have, however, a very limited time frame --- typically a few years or decades---, and mainly concerns the contemporary period \cite{batty2013new,barthelemy2016structure,bettencourt2021introduction}. Other studies that consider historical periods, such as US counties between 1840 and 1990 \cite{beeson_population_2001}, focus on the simpler quantity that is population, and other aspects such as the location and number of different activities are rarely considered on an historical scale. The construction of a science of cities however needs a long time span, and a quantitative approach to the historical evolution of urban systems is key for our understanding of these systems. Cross-sectional studies are available such as in \cite{ortman_pre-history_2014,ortman_cities_2020,lobo_settlement_2020}, but an analysis on individual cities over a long time is very demanding in terms of dataset construction. This might however change as we witness a spectacular increase of the digitalization of historical sources, which opens the way to a broader view on the evolution of urban systems. 

Existing studies consist essentially on the digitalization of old maps allowing scientists to characterize the evolution of the road system \cite{perret2015roads, el_gouj_urban_2022}, a critical component in cities. Very few other aspects were considered, with the exception of urban forms as in \cite{bretagnolle_urbanization_2015, burghardt_analyzing_2023}. However, it is usually advocated that economic activities lie at the heart of a city allowing for its specialization and diversification through time \cite{rodriguez_urban_1986}. An important question is then to understand and to identify economic activities that go along the growth of a large city. Such data is however extremely difficult to get for historical periods and here we take advantage of a new dataset recently released \cite{geohistoricaldata_annuaires_2023}, constructed from the city directories that consist of large lists of individuals, merchants or prominent inhabitants, businesses, organizations and institutions. We report the analysis of such a dataset obtained for the city of Paris in the period 1829-1907 which gives access to a total of about 1 million entries of economic activities with their address, regularly updated in this period  (see details about the dataset construction in the Methods section). This $79$ years period allows us to have an historical perspective on city growth and activity evolution, but also to question the impact of large perturbations that took place in this period, such as Haussmann's works that happened between 1853-1870 and modified in-depth the structure of the city \cite{barthelemy_2013}, or polical conflicts such as the Franco-Prussian war and the Commune `revolution' (1870-1871).


During the evolution of a city we observe the emergence of different activities and their variation as the city grows. The idea here is then to characterize the different activities and to exhibit the existence of categories of activities that have different functions. A natural tool for the analysis of the evolution of activities in a city is scaling \cite{pumain2004scaling, bettencourt_growth_2007}. It has indeed been observed that a total quantity $Y$  measured for a city of population $P$ varies as $Y \sim P^\beta$, with $\beta>0$. In other words, the quantity per capita $Y/P$ varies as $P^{\beta-1}$ which allows us to identify three different categories. First, for $\beta=1$, the quantity per capita is constant. This typically corresponds to quantity that are relative to human needs (such as water consumption). In the other cases, the quantity per capita depends on the city size. For $\beta>1$, there is a `positive' impact of the city: the larger it is and the larger the quantity $Y/P$. In contrast, the case $\beta<1$ usually corresponds to economy of scale. These scalings are usually measured `cross-sectionally' at a given time for different cities with different values of their population. The number of studies considering history and evolution through time are increasing (see Table~\ref{tab:litteraturereview}), but the majority concerns cross-sectional scaling analyses for a given period, and we identified two historical studies of cross-sectional scaling over time only: one focusing on election results in the US between 1948 and 2016 \cite{bokanyi_universal_2018}, the other one on the morphological structure of US cities between 1900 and 2015 \cite{burghardt_analyzing_2023}. For contemporary cities, studies are mainly based on cross-sectional scaling over time (computed over various cities at a given time and also for different dates), and it is not yet clear how this relates to temporal scaling (computed over time for a given city) \cite{depersin_global_2018, keuschnigg_scaling_2019,bettencourt_interpretation_2020}. Indeed, analyzing cross-sectional data differs from studying the temporal scaling of individual cities. The former primarily reflects spatial dynamics, enabling insights into the characteristics of different cities at specific time points, while the latter, focusing on the temporal dynamics, offers a view into how individual cities evolve over time. Here, we won't address this general problem, but will focus on the measured value of $\beta$ as a way to categorize a wide range of urban activities and to characterize the evolution of a city. The objective is then to analyse how the number of a given activity grows when the population varies.

\begin{table}
	\begin{center}
		\caption{Scalings in historical and temporal studies of spatial systems. CS refers to the number of case studies and the quantity $\Bar{Q}$ to the average number of different quantities analyzed in case studies. Data for the literature review is presented in the section 1 of SI.}
		\resizebox{0.95\width}{!}{
		\begin{tabular}{lrrrr}\label{tab:litteraturereview}
			Period begins  & Scaling type  & CS & $\Bar{Q}$ & Reference \\
			\midrule
			\multirow{2}{*}{after 1960}  & cross-sectional over time  & 21   & 4.95   & \cite{pumain_evolutionary_2006, ignazzi_scaling_2014, conde_mutations_2015,paulus_knowledge_2016, meirelles_evolution_2018,goncalves_scaling_2014,depersin_global_2018, keuschnigg_scaling_2019,xu_cross-sectional_2019, bettencourt_interpretation_2020, hong_universal_2020,ribeiro_relation_2020, xu_scaling_2020,finance_scaling_2020, maisonobe_regional_2020, andersson_geography_2020, arvidsson_urban_2023,lei_urban_2022, curado_scaling_2021, strumsky_as_2021} \\
			& temporal     & 8     & 4.5  & 
			 \cite{goncalves_scaling_2014, depersin_global_2018, keuschnigg_scaling_2019,xu_cross-sectional_2019, bettencourt_interpretation_2020, hong_universal_2020, ribeiro_relation_2020, xu_scaling_2020} \\
			\multirow{2}{*}{before 1960} & cross-sectional one period & 17  & 1.94  & \cite{ortman_settlement_2016, cesaretti_populationarea_2016, ortman_why_2020, hanson_urbanism_2017, hanson_urban_2019, altaweel_urban_2019, cesaretti_increasing_2020, smith_lowdensity_2021} \\
			& cross-sectional over time  & 2   & 3  &
			 \cite{bokanyi_universal_2018, burghardt_analyzing_2023} \\
			\bottomrule
		\end{tabular}
	}
	\end{center}
\end{table}

\section*{Results}

\subsection*{Overview of Paris Activities}

The dataset studied here encompasses about 1 million entries documenting the primary activities within Paris. Each entry corresponds to an individual having an activity (and less frequently to an organization), together with an address. These entries display a consistent growth pattern, expanding from 23,000 in 1829 to 88,000 in 1907. In order to compare historical activities of Paris during the 19\textsuperscript{th} century with contemporary cities, we have used 21 contemporary groups, drawing inspiration from the North American Industry Classification System (NAICS) developed by the U.S. Census Bureau (more details in the Methods section).

\begin{figure}
	\centering
	\includegraphics[width=\linewidth]{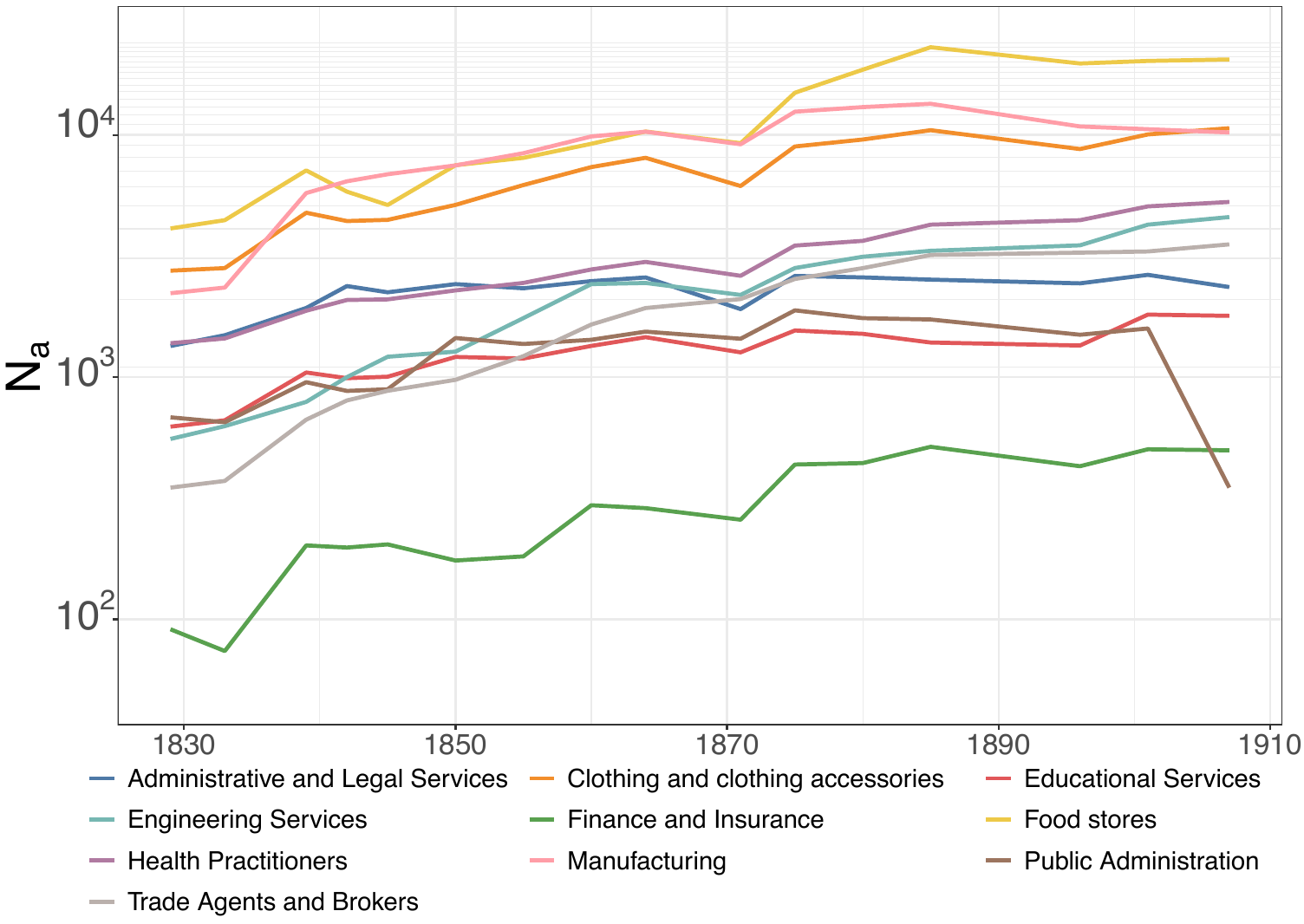}
	\caption{Number of city directory entries for different activity categories ($N_a$) over time. These activities all grow on average but at different rates. We show here a sample of activity categories and the results for all of them can be found in Fig. S1 of the SI.}
	\label{fig:Na}
\end{figure}

We show the evolution in time of a selection of activity categories in Fig.\ref{fig:Na}.
We observe that there is overall increase in activity categories -- with the exception of the decline in the public administration sector at the end of the period, resulting from a bias in the dataset (see details in Material section). The two most significant sectors correspond to food stores and clothing related activities. The number of food stores increased from $4,000$ in 1829 to $20,000$ in 1907, and for clothing from approximately $3,000$ entries in 1829 to $10,000$ in 1907. Manufacturing appears as the third most important section with over $10,000$ entries by 1907. The Gini coefficient --- that characterizes the heterogeneity --- computed on the number of entries $N_a$ for 21 categories is relatively stable and fluctuates between $0.44$ and $0.52$ showing that the global inequality of activity sizes is stable over time (SI, Fig. S2). Despite this stability, the increase rate of $N_a$ differs largely between categories. Notably, there is a tenfold increase in the number of trade agents and brokers between 1829 and 1907, along with an eight-fold increase in engineering and more than fivefold for finance and insurance individuals. Additionally, administrative and legal individuals, along with educational ones, displays  a modest increase of 1.75 and 3 times, respectively.

\subsection*{Scaling of Activities and Population over Time}

\begin{figure}
	\includegraphics[width=\linewidth]{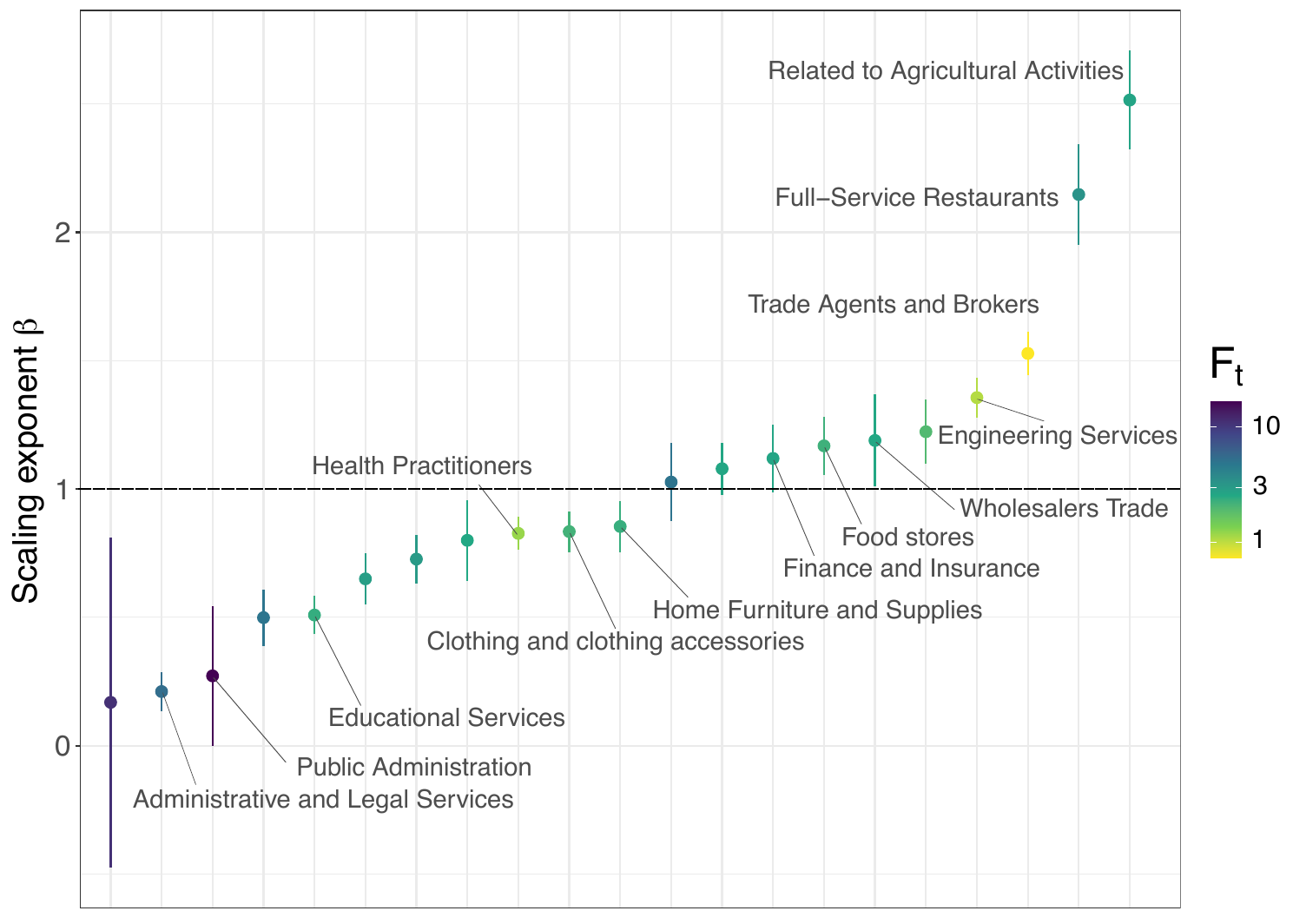}
	\caption{Exponent $\beta$ for different activity categories. The power-law fits in log-log representations are shown in SI Fig. S3 and the values of $\beta$ and standard errors are shown in Tab. S3. 15 points in time were computed for each category since population censuses were carried out approximately every 5 years). The quantity $F_t$ is the fluctuation measurement of the slope for $N_a$ relative to the average slope  ($\bar{S}_a$), and given in SI Tab. S4 for all categories.}
	\label{fig:scaling}
\end{figure}

The population of Paris increased strongly from 785k in 1831 to more than 2.75M in 1907 \cite{cristofoli_populations_2023}, questioning how activities in the city  correspondingly grew. The very sharp growth in the population of the municipality of Paris is closely linked to the process of urban sprawl that led to the redrawing of the city's administrative boundaries in 1860 (see Methods). A natural tool \cite{depersin_global_2018} to investigate this question is to analyze how the number of entries $N_a(t)$ for a given activity `$a$' scales with the population $P(t)$ for all times in our dataset. We expect in general a power-law of the form 
\begin{align}
	N_a(t) \sim P(t)^\beta
\end{align}
where the exponent $\beta$ characterizes the growth rate of the activity `a'. We measure this exponent for different categories of activities and obtain the result showed in Fig.\ref{fig:scaling}.
We observe that these scaling exponents are highly variable, ranging from $\beta \approx 0.21$ for administrative and legal category to $\beta \approx 2.52$ for activities related to agriculture. A value of $\beta \sim 1$ indicates that an activity grows linearly with population during the period. We observe such a linear relation for building and car rental categories ($\beta \approx 1.03 \pm 0.15$) and non-food retailers (except for clothing and home furniture sales). Other activities display a value of $\beta$ slightly below 1, such as clothing and home supplies manufacturers-retailers (respectively $\beta \approx$ 0.84 and 0.85), or health practitioners ($\beta \approx$ 0.83) indicating a growth rate slightly lower than the population. We also observe $\beta$ values slightly larger than one for food stores ($\beta \approx 1.17 \pm 0.11$), although a linear behavior cannot be excluded. We note that food stores described in the directories display highly variable values of $\beta$ depending on the type of food (see Fig. S4 and S5 in the SI). Grocers grow linearly with population ($\beta \approx 0.95 \pm 0.05$), whereas we observe for seed or cheese retailers an almost constant behavior ($\beta \approx 0.04$ and $0.07$, respectively). 

Besides these activities that scale approximately with the population, we observe activities with scaling exponents very different from 1. For example confectioners 
($\beta \approx 0.59$), wine retailers ($\beta \approx$ 1.48) and creamers ($\beta \approx$ 2.38). Furthermore, activity categories relative to politico-institutional services, such as administrative and legal individuals, public administration or educational individuals grow sublinearly to the population, with $0.21 < \beta < 0.51$. In contrast, engineering individuals ($\beta \approx 1.36$), trade agents and brokers ($\beta \approx 1.53$), and full-service restaurants ($\beta \approx 2.15$) experience more rapid growth rates than Paris' population during the period.

A wide range of empirical studies on urban scaling analyse data of system of cities by comparing different cities at a given time. Among these studies, Pumain et al. \cite{pumain_evolutionary_2006} compare $\beta$ values for various economic and social activities within contemporary city systems in both the US and France (in 2000 and 1999). Certain exponents identified in their study, such as finance and insurance, and wholesalers trade exhibit similarities to the scaling observed here for Paris during the 19\textsuperscript{th} century (see Table~\ref{tab:beta}).  It seems that in these cases, the value for $\beta$ is governed by principles independent from the period. Moreover, comparing to the Brazilian system of cities in 2000, the wholesale trade sector had a $\beta$ value of 1.18 \cite{ignazzi_coevolution_2015}. For other activities, the $\beta$ values differ significantly. This is particularly true for the cases of education, art, construction and restaurants, where values can be larger or smaller than $1$ depending on the measure, and whose evolution depends strongly on the period considered. It should be noted that $\beta$ values can vary in space and time, as in the case of educational services: indeed, we observe both a variation in space ($\beta$ for the French case is very different from that of the US) and in time ($\beta$ for Paris in the 19\textsuperscript{th} century is very different from that of France today). This suggests that the total number of workers in education services depends strongly on other considerations such as political decisions. In addition, while some values are similar between US and French systems, as in the construction sector, they may be different in other emerging economies. Indeed, $\beta$ value of this sector is about 1.08 in China (2000) and 0.98 ten years later \cite{finance_scaling_2020}, while they are 1.13 (2000) and 1.32 (2010) in Brazil (due to the housing boom and population concentration in larger cities \cite{ignazzi_coevolution_2015}).

\begin{table}
	\begin{center}
		\caption{Comparison of $\beta$ scaling. The first column (Paris) shows the results obtained here, and in the other columns we show the results obtained for the US and France in \cite{pumain_evolutionary_2006,um_scaling_2009}.}
		\resizebox{0.85\width}{!}{
		\begin{tabular}{lr|rr}\label{tab:beta}
			& temporal $\beta$ & \multicolumn{2}{l}{cross-sectional $\beta$} \\ \midrule
			Activity Category & Paris & U.S. & France  \\
			\midrule
			Finance and Insurance & 1.12 & 1.22 & 1.19 \\
			Wholesalers Trade & 1.19 & 1.19 & 1.07 \\
			Health practitioners & 0.83 & 0.95 & 0.99 \\
			Educational services & 0.51 & 1.23 & 0.97\\
			Arts, Entertainment and Recreation & 0.73 & 1.14 & 1.10\\
			Construction & 0.50 & 1.05 & 0.98 \\
			Food services & 1.17 & 0.98 & 0.99 \\
			Restaurants & 2.15 & 0.93 & \textit{NA} \\
			\bottomrule
		\end{tabular}
	}
	\end{center}
\end{table}

Despite these fluctuations in time and from a country to another, these values allow us to distinguish the role of different activities during the growth of a city. We thus propose the following typology of activities for the dynamics of urban activities. For $\beta<1$, we find all institutional services whose number is in general dictated by optimization principles based on accessibility \cite{stephan1977territorial, gusein1982bunge, gastner2006optimal,um_scaling_2009}. Typically, minimizing the total time expenditure for public facility leads to a power law behavior with $\beta<1$. The underlying assumption being that social structures evolve in such a way to minimize the time spent in their operation (see Methods for an analytical discussion). It is important to note that entries in directories relating to institutional services are not limited to establishments (e.g. schools for educational services), but also to individuals (e.g. teachers), as presented in SI Table S1. In this respect, optimization is also driven by other factors in addition to accessibility. Urban activities with $\beta\approx 1$ are intrinsically linked to population and to a fixed number per capita. The number of these activities is mainly driven by the population needs (typically as it is the case for grocers). These two activity types ($\beta<1$ and $\beta\approx 1$) correspond essentially to basic needs and can be expected for other growing modern cities. This is in contrast with the last type of activities that display $\beta>1$ whose number is mainly dictated by specific needs, that in general will vary according to the period considered. Some activities depend on transformations in the economic production system, like engineering services during the industrial revolution, while others depend on transformations in consumption habits. This is typically the case here of wine retailers (owing to the increasing popularity of wine consumption in the 19\textsuperscript{th} century \cite{nourrisson_7_2017}) or creamers (milk consumption increased during this period \cite{husson_consommation_1875}). These  large $\beta$ values reveal a specific phase of the development of the city: urban sprawl (e.g. the number of individuals in charge of feeding cows for milk production and trade increase because the proximity to consumers is vital before pasteurization methods were invented), or a specialization (such as restaurants in Paris).

\subsection*{Dynamics of Activities}

\begin{figure*}
	\centering
	\includegraphics[width=17.8cm,height=7.13cm]{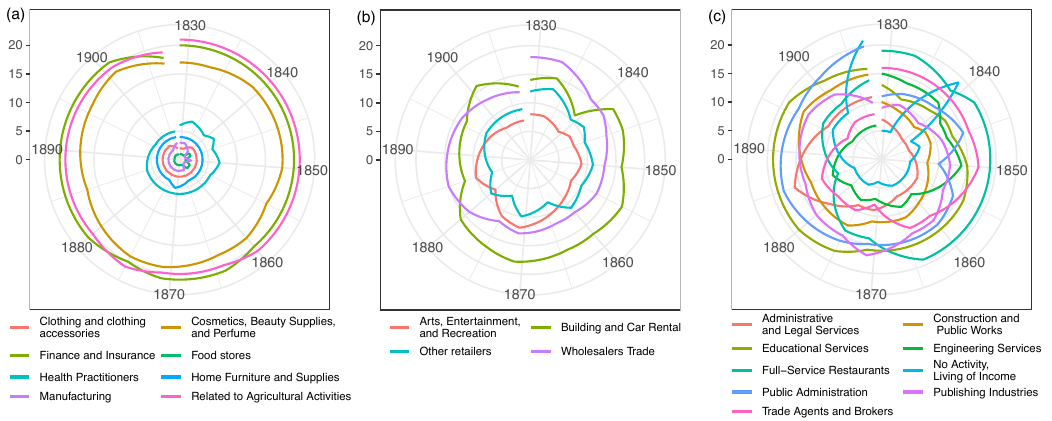}
	\caption{Rank clock for activity categories. The rank $r$ of the number $N_a$ of directory entries related to a given category is calculated at each time step. The radius is increasing with the rank, and the most important categories (small ranks) are then represented by small circles. We represent here activities in three panels according to the temporal behavior of their rank. (a) We show categories with minor rank changes (that have gained or lost at most 3 ranks between 1829 and 1907). (b) Categories that have gained or lost from 4 to 7 ranks, and (c) categories that experienced major rank changes (more than 7 ranks during the period considered here).}
	\label{fig:rank-clock}
\end{figure*}

In addition to the scaling, we also investigate the dynamics of activities. An interesting visualization tool is the rank clock \cite{batty_rank_2006} for activity categories that we show in Fig.\ref{fig:rank-clock} for the period 1829-1907, 
and observe different types of dynamics. First (Fig.\ref{fig:rank-clock},a), there are activities -- such as food stores or others such as cosmetics, beauty supplies or perfume -- that maintain their rank over 79 years. More generally, all activities connected to everyday needs of residents (food stores, clothing manufacturers/retailers, or heath practitioners, etc.) have a rank that remain stable during the period under study. Second, there are activities whose rank decreased (i.e. became more important) such as wholesalers trade, trade agents and brokers, restaurants and engineering individuals (Fig.\ref{fig:rank-clock},b,c). The latter had its important sharp increase, moving from the 15\textsuperscript{th} position in 1829 to the 6\textsuperscript{th} position in 1907. Finally, there are activities whose rank increased such as administrative and legal, educational, public administration, construction and public works, and publishing industries ones. Their number (see Fig.\ref{fig:Na}), however, remained stable over time which is consistent with the fact that it is governed by optimal considerations.

\subsection*{Effect of large perturbations}

Paris during the 19\textsuperscript{th} century experienced large perturbations. One of the most important happened under the guidance of Baron Haussmann \cite{jordan_transforming_1995,samuels_urban_2004,barthelemy_2013}. The social, political, and urbanistic importance and impact of Haussmann's renovation is particularly significant. Essentially, until the middle of the 19\textsuperscript{th} century, central Paris has a medieval structure composed of many small and crowded streets, creating congestion and, according to some contemporaries, probably health problems. In 1852, Napoleon III commissioned Haussmann to modernize Paris by building safer streets, large avenues connected to the new train stations, central or symbolic squares (such as the famous place de l’Étoile, place de la Nation and place du Panthéon), improving the traffic flow and, last but not least, the circulation of army troops. Haussmann also built modern housing with uniform building heights, new water supply and sewer systems, new bridges, etc. These works lasted between 1853 and 1870.

The second important historic event is called the Paris Commune which was a revolutionary government that seized power in Paris in 1871. It emerged after the defeat of France during the Franco-Prussian war in 1870-1871 (with the siege of Paris by Prussian since September until the capitulation of Paris), and in opposition to the new government formed by the monarchist-majority National Assembly in 1871. Paris Commune government lasted from March to May, while the regular government troops once again surrounded Paris, until the `Communards' were defeated in a severe crackdown.

An interesting question is then whether we can observe these two different types of events in the statistics of activities. Indeed, Haussmann's project was an extensive urban planning operation lasting 17 years, while the Paris Commune was a political event lasting just a few months. We thus compute the average slope for a given activity (computed over the period 1829-1907), and the average relative deviation $D_a$ around this slope computed over the Haussmann period (1855-1875), during the Paris Commune (1864-1871), and after the Paris Commune (1871-1875). More precisely, if we denote by $S_a^{loc}$ the local slope for the activity `a' (equal to $(N_a(1875)-N_a(1855))/20$ for the Haussman period for example), and by $\overline{S}_a$ the average slope for activity `a', the quantity $D_a$ is given by
\begin{align}
	D_a=\frac{S_a^{loc}-\overline{S}_a}{|\overline{S}_a|}
\end{align}
The results for different activities are shown in the Table~\ref{tab:slope} and display large impacts on some activities (see also SI Fig. S6 and S7).

The results for different activities are shown in the Table~\ref{tab:slope} and display large impacts on some activities.
\begin{table}[h!]
  \caption{Mean slope and relative deviations of activity categories considering different perturbations in the history of Paris. The quantity $\bar{S}_a$ is the mean slope computed over $1829-1907$ and $D_a$ is the relative slope deviation computed for: a) the Haussmann period (1855-1875); b) the Paris Commune (1864-1871); c) after the Paris Commune (1871-1875). The results shown in \textit{italic} correspond to calculation based on the mean slope time period reference 1829-1901, i.e. without the source effects between 1901 and 1907 (see Methods section). The specific cases of ‘No activity, living of income’ and ‘Public administration’ (denoted by *) displays a highly exceptional pattern and are discussed in the Methods section. }
	\resizebox{0.85\width}{!}{
	\begin{tabular}{lrrrr}
		\label{tab:slope}
		Activity Category  & $\bar{S}_a$ & $D_a$ (a) & $D_a$ (b) & $D_a$ (c) \\
		\midrule
		Administrative and Legal Services  & 12.96    & 0.12   & -8.41  & 12.67\\
		Arts, Entertainment, and Recreation    & 30.22  & -0.16 & -4.21  & 5.86\\
		Building and Car Rental  & 18.10 & 1.29  & -4.62 & 10.64 \\
		Clothing and clothing accessories & 101.58   & 0.36   & -3.67  & 5.92 \\
		Construction and Public Works & 13.78  & 1.39 & -7.68 & 12.86 \\
		\begin{tabular}[c]{@{}l@{}}Cosmetics, Beauty Supplies, \\ and Perfume Retailers\end{tabular}  & 17.74   & -0.22   & -3.31 & 4.34  \\
		Educational Services & 14.97  & 0.22  & -2.88   & 3.89 \\
		Engineering Services  & 51.68  & 0.04 & -1.73 & 2.09 \\
		Finance and Insurance   & 5.23  & 1.43 & -1.82  & 7.51   \\
		Food stores  & 210.42  & 0.64 & -1.74  & 5.77 \\
		Full-Service Restaurants & 22.76    & 2.41   & 1.44  & 5.34 \\
		Health Practitioners  & 50.17  & 0.04  & -2.06  & 3.36 \\
		Home Furniture and Supplies   & 79.4    & -0.17  & -4.88 & 6.44\\
		Manufacturing   & 102.99 & 0.98 & -2.61 & 7.08 \\
		No Activity, Living of Income* & \begin{tabular}[c]{@{}l@{}} -23.09 \\ \textit{13.6} \end{tabular} & \begin{tabular}[c]{@{}l@{}}0.96 \\ \textit{-1.07} \end{tabular} & \begin{tabular}[c]{@{}l@{}} -3.5 \\ \textit{-8.65} \end{tabular} & \begin{tabular}[c]{@{}l@{}}7.46 \\ \textit{9.98} \end{tabular} \\
		Other retailers & 33.99 & 0.75 & -4.26 & 8.2 \\
		Public Administration*  & \begin{tabular}[c]{@{}l@{}}-4.27\\ \textit{12.6} \end{tabular} & \begin{tabular}[c]{@{}l@{}} 7.06 \\ \textit{1.05} \end{tabular}  & \begin{tabular}[c]{@{}l@{}} -2.38 \\ \textit{-2.15} \end{tabular} & \begin{tabular}[c]{@{}l@{}}27.18 \\ \textit{7.87} \end{tabular} \\
		Publishing Industries & 21.06 & 0.25 & -4.34 & 8.19 \\
		Related to Agricultural Activities & 5.01 & 1.93 & -1.46 & 4.88\\
		Trade Agents and Brokers & 40.72  & 0.62 & -0.4 & 1.71 \\
		Wholesalers Trade  & 24.55 & 1.21 & -3.83  & 7.09 \\ 
		\bottomrule
	\end{tabular}
}
\end{table}

During the Haussmann period, most activities were positively affected with an increase of their numbers. In particular, construction works and building rental are impacted, as could be expected for such a large-scale restructuring of the city and its infrastructures. Indeed, during the beginning of Haussmann's renovation we observe the growth of new companies in the construction sector ($D_a \approx 3.34$ in 1855-1864 and $-0.21$ in 1864-1875), while furnished hotels (building rental) used as transitional housing for new immigrants to Paris are mainly impacted during the 1860s ($D_a \approx 0.83$ in 1855-1860; $2.87$ in 1860-1864), probably as a consequence of the housing price inflation of over $50\%$ in the 1850s in the city center \cite{faure_speculation_2004}. From this perspective, the global impact of this large-scale planning operation was positive. It is very different in the case of the Paris Commune that had a globally negative impact on most activities, as we could expect for such a troubled period. The impact is not significant when looking at the total number of entries in city directories ($N=93.7k$ in 1864, $101.8k$ in 1871 and $107.5k$ in 1875), but it is significative on the most frequent activities in the city analyzed here (respectively $N=67.9k$, $56k$ and $80.8k$). Surprisingly enough, the period after this event witnessed a large-scale reorganization with many activities whose number exploded. From this point of view, three main factors probably explain this evolution: 1) temporary closures of companies; 2) temporary transformations in the types of activity carried out by companies listed before, during and after 1871; 3) changes in the ways in which individuals in 1871, especially elites, present themselves in the directory (e.g. administrative and legal individuals, or by people living off their income as owners or annuitants). Despite the political and social shock of the two sieges of Paris in 1870-1871, the transformations of activities are likely to be short-lived rather than long-lasting. A return to regular activity can be observed from 1875, in terms of the number of activities by category (see SI Figs. S6). 


\section*{Discussion}

We showed on the case of Paris during the 19\textsuperscript{th} century that not all activities are equivalent and that they can be grouped in categories having different dynamics and scaling. In particular, it seems that some activities naturally accompany the growth of the city and can be considered as intrinsic to its development. These activities essentially answer to basic needs of the residents, and the type of scaling with population of the number of these activities depends essentially on their underlying logic: it seems indeed intuitive that the number of food stores scales more or less linearly with population while the number of administrative or educational needs follow an optimization logic (and scale sublinearly). All these activities constitute the core of the city, while other activities determine its specific features. These quickly developing activities usually respond to some passing fad or to a need of the specific period and the phase of the city's development. 

The dynamics of activities in the city over time should also be considered from the perspective that Paris is a city within a system of cities \cite{berry_cities_1964}, and is also the largest city in the French urban system. In this perspective, the evolutionary theory for interpreting urban scaling laws \cite{pumain_evolutionary_2006} can shed some light on our results. Apart from human-related needs that are always characterized by $\beta\approx 1$, we can discuss the typology in terms of innovation cycles. When the exponents are larger than $1$, the corresponding activities are in a first adoption stage of innovations in the urban systems, and are usually concentrated in the largest cities, while exponents below $1$ correspond to more mature activities \cite{pumain2021networks}. We thus expect the typology of these activities to be valid for all growing cities, with differences being essentially in the 2nd group of activities ($\beta>1$) that characterize the city's specific features in an innovation phase or specialization. These results shed a new light about the development of large cities by highlighting the existence of a typology of activities exhibiting their different role that they play in the development of a large city.



\subsection*{Data availability} 

The dataset of the population censuses of Paris at the scale of districts between 1801 and 1911 is openly accessible with the documentation on Nakala platform of the CNRS Research Infrastructure Huma-Num  \href{https://doi.org/10.34847/nkl.e173c93p}{DOI:10.34847/nkl.e173c93p}. The dataset of Paris directories entries with NAICS inspired categories between 1829 and 1907 specifically constructed and used for this paper is openly accessible on Zenodo platform \href{https://doi.org/10.5281/zenodo.8388101}{DOI:10.5281/zenodo.8388101}.

\subsection*{Code availability}

The open repository \href{https://doi.org/10.5281/zenodo.8388101}{DOI:10.5281/zenodo.8388101} contains the code to create the figures and tables of both the main text and the SI.

\subsection*{Acknowledgements}
We thank N. Abadie, S. Baciocchi, P. Cristofoli, B. Duménieu, E. Carlinet, J. Chazalon and J. Perret, who set up the processing chain enabling the extraction and enrichment of Paris directories, and without whom this work would not have been possible. This research was funded by ANR SoDUCo Program, grant number ANR-18-CE38-0013 (JG and MB).

\subsection*{Author Contributions Statement}
JG and MB designed the study, JG worked on the data and analyzed it. JG and MB wrote
the paper.

\subsection*{Competing Interests Statement}
None.

\clearpage
\newpage

\section*{Materials and methods}
\subsection*{Paris Directories between 1799 and 1914}

City directories became an international publishing phenomenon during the 19\textsuperscript{th} century in Europe and USA. They consist of large lists of individuals, merchants or prominent inhabitants, businesses, organizations and institutions, each with description and address. Published at a fast pace, often yearly, they provide massive, fine-grained and highly valuable geohistorical information for in-depth interdisciplinary studies of social and spatial aspects of cities. Analyzing the content of these directories over time and with a high temporal frequency requires a considerable amount of manual transcription, structuring, geolocation and data linking work. To overcome this difficulty, SoDUCo ANR Program has proposed an automatic pipeline to extract, semantically annotate, geocode and analyze the contents of 141 directories of Paris published between 1799 and 1914. First, each page is processed using image segmentation to extract its layout and detect directories entries, i.e. regions containing a triple composed of a person (physical or moral), an activity and an address. OCR is applied on each entry to get its textual content, which is then semantically enriched using a deep-learning based Named Entity Recognition approach \cite{abadie_benchmark_2022}. Finally, each address is assigned a geographic position in the city using a geocoding process. Spatial enrichment leverages historical maps of Paris so the geocoding takes into account the changes in the numbering and streets over the century, including Haussmann’s renovation \cite{cura_historical_2018}. More than 10M entries currently compose the 141 digitally transcribed and enriched directories \cite[version 3, alphabetical lists]{geohistoricaldata_annuaires_2023}.

Paris directories are commercial editions, involving competition between publishers, buy-outs over time and periods of editorial monopoly. Five main phases have been revealed throught a systematic chronological inventory of directories. The years 1780-1793 are those of the origins. It was followed by the advent of the \textit{Almanach du commerce} (1798-1815). Competition was fierce, and publications abounded until 1856, when the Firmin Didot brothers bought the Bottin publishing company. The period 1857-1890 was thus marked by the hegemony of the Didot-Bottin collection. A new period of competition began in 1891.

The city of Paris is conceived by the editors in terms of its socio-economic functioning. The main issue is to list individuals so that other ones can learn about their activities and addresses, and thus potentially interact with them (through visits or epistolary exchanges). The aim is to connect individuals in the city and from the city with others who are part of wider networks by establishing these lists; even including, if necessary but very rarely, individuals who live or perform their activities outside the administrative limits of the city, but belong to socio-economic networks of Paris (as in the case of Batignolles before 1860, see SI section 6, Fig. S10). In this sense, directories are homogenous across space. Moreover, activities listed in the directories are limited to those associated with specific addresses, implying a focus on activities linked to the physical structure of the city. Consequently, informal activities are not listed in the directories, and the data captured therein mirrors the evolution of the city's urban fabric. In this sense, the coverage of the city's outskirts can be less extensive than that of the center, since it is also on some of these fringes that informal activities are more numerous.

The lists in directories follow three main organizations: by name, by profession, and by street, in order to simplify researches for readers of the time. Payment is not required to be listed in the directories. However, the differentiated length of activity descriptions in the professional lists, and the integration of promotional vignettes towards the end of the 19\textsuperscript{th} century, testify to the payment of advertising by companies. In contrast, the alphabetical lists studied in this paper are made up of much shorter, standardized activity descriptions.

\subsection*{Selection of Directories and Main Activities between 1829 and 1907}

The fierce competition between editors from the mid-1810s onwards led them to enrich the listings by extending their social and economic coverage, and to continually update their editions. Indeed, until 1817, alphabetical lists of Paris directories exclusively documented prominent inhabitants. Subsequently, a diverse range of individuals was incorporated into the alphabetical lists, enabling the study of the dynamics of diversified urban activities. However, the statistical distributions of the number of entries per page in alphabetical lists show significant variations for each of the directories published before 1829, resulting from noise in the segmentation phase of the digitized directory images. Hence, we took into account the data from 1829 onwards.

Knowing that the directories capture activities related to the urban physical space, we accordingly selected directories with a time step of around 5 years. Indeed, changes in urban structure do not occur at an annual rate. The choice of a particular year over another was determined through an evaluation of the extraction and enrichment quality of the digital transcriptions of the directories.

The character strings of the activities in the data are rather noisy, due to OCR errors, and to the activity description in the sources (use of abbreviations and/or information add-on). We thus applied the soundex phonetic index to smooth out the noise and group together similar activities. Over 7,000 different activities are listed in the directories studied. The rank-size distributions of the activities show that a few activities group together a large number of persons, such as wine retailers in 1885 counting 10,000 entries, while a large number of activities group together a small number of persons (see SI Fig. S8, S9). We selected the 175 most frequent activities from the slightly less than 1.4M entries in the 16 directories studied, which account for 71\% of all entries, i.e. over 990,000 entries.

\subsection*{Activity Categories Inspired of the U.S. Census Bureau's NAICS}

The 175 most frequent activities are grouped in categories inspired of the 2022 North American Industry Classification System (\href{https://www.census.gov/naics/}{NAICS}). The latter is a standard, and facilitates comparisons with present-day cities. The social and economic organization of cities during the 19\textsuperscript{th} century and nowadays is not strictly equivalent, so we could not systematically apply NAICS categories on past-activities described in directories. As an example, places of manufacturing and retail are regularly not distinct, requiring their association in some of the categories proposed (see SI Tab. S1). We studied the definitions of each activity description with four editions of the dictionary of the French Academy (\textit{Dictionnaire de l'Académie française} from 1798 to 1935) to better understand what covers each of them and to evaluate semantic stability/instability of the words used in directories. This method step was necessary to relate with accuracy the activity descriptions in directories with NAICS-inspired categories typology.

Soundex index is not always efficient when OCR errors in the data are very important at the begin of character strings of the activities, and sometimes groups very different activities that sound alike. We thus performed manual editing of data to reduce noise of bad automatic attribution between activities in Paris directories and activity categories. Some variations in the number of entries are significant, as in the case of food stores, where total entries were 250,000 before the manual revision and 175,000 after (see SI Fig. S9).

\subsection*{The specific cases of `No activity, living of income' and `Public administration'}

The category `No activity, living of income' displays a highly exceptional pattern, characterized by a significant decline of $N_a$ in both 1839 and 1907 (see SI Fig. S6). It reflects editorial choices made by the companies that published directories. Indeed, this category encompasses individuals who, in 1839, were classified not as owners by publisher Lamy, but rather as electors or eligible individuals, within the context of a tax-based voting system -- payed by approximately 1.2\% of Parisians at that time and about 0.65\% of French population in 1842 \cite{tudesq_1958}. Furthermore, starting from 1903, the publishers Didot-Bottin undertook the creation of a distinct directory dedicated to prominent inhabitants (the \textit{Bottin mondain}). As a result, all 2,780 individuals listed as owners or annuitants in the 1901 directory are absent in the 1907 edition. This editorial shift also explained the decrease of $N_a$ in public administration category (Fig.\ref{fig:Na}). Specifically, 1,240 individuals were removed from 1901 to 1907 directories, highlighting that 78\% of individuals listed in this category were prominent inhabitant in Paris (e.g. deputies, ministers, advisors at the Court of cassation). In contrast, certain other activities, such as finance and insurance individuals or health practitioners, demonstrated a consistent presence over time, suggesting that bankers, placement agents, doctors, etc. were not part of the highest social strata of society according to the publishers of the time.

\subsection*{Relationships between administrative and urban activities delineations of Paris}
The question of Paris' delineation is an important one, as it generally affects scaling results \cite{cottineau_diverse_2016}. Paris directories are not based on the municipality's administrative boundaries, but reflect the economic activities of the urban area. In this sense, the spatial spread of geolocalized entries over the century reflects urban sprawl -- mainly from the 1870's. Population censuses, on the other hand, are carried out within the districts of the municipality of Paris. In this sense, the delineation is administrative. In addition, the political and fiscal law of June 16, 1859 implied the enlargement of the municipality of Paris by merging and redrawing the boundaries of other neighbouring municipalities \cite{montel_lagrandissement_2012}. The areas covered by the activities contained in the directories are however overwhelmingly included within the city's administrative boundaries over time (see SI section 6, Fig. S10).\\

\subsection*{Optimal number of public facilities}

Elaborating on a argument proposed by Stephan \cite{stephan1977territorial}, Um et al \cite{um_scaling_2009} minimize the total time expenditure to explain the scaling with population. We assume that the city has a surface area $A$ and contains $N$ facilities. 
The total time expenditure can be written as
\begin{align}
	T=hN+cPd^\gamma
\end{align}
where $d=\sqrt{A/N}$. In the first term of the r.h.s. the quantity $h$ represents the number
of person.hour per establishment, and $hN$ is then the total person.hour expenditure for
the $N$ facilities. The second term corresponds to the time needed for the population $P$ to reach the facility of the form
$d/v$ where $v$ is the average velocity. The dependence on distance could however be  more complex
and we generalize this expression and write $d^\gamma$ where $\gamma$ is an unknown exponent. 

Minimizing $T$ with respect to $N$ then gives
\begin{align}
	N\sim A^{\gamma/\gamma+2}P^{2/\gamma+2}
\end{align} 
leading to an exponent $\beta=\frac{2}{\gamma+2}$. The usual assumption $\gamma=1$ leads
to the standard $\beta=2/3$ as discussed in  \cite{gusein1982bunge,gastner2006optimal}.


%

\end{document}